\documentclass[%
prd
 ,twocolumn%
 ,secnumarabic%
,amssymb, amsmath,nobibnotes,showpacs]{revtex4}
\usepackage{bm}%
\expandafter\ifx\csname package@font\endcsname\relax\else
 \expandafter\expandafter
 \expandafter\usepackage
 \expandafter\expandafter
 \expandafter{\csname package@font\endcsname}%
\fi

\begin{document}
\title{Implications of the gauge-fixing in Loop Quantum Cosmology.}%

\author{Francesco Cianfrani$^{1}$, Giovanni Montani$^{123}$}%
\email{montani@icra.it, francesco.cianfrani@icra.it}
\affiliation{$^1$Dipartimento di Fisica, Universit\`a di Roma ``Sapienza'', Piazzale Aldo Moro 5, 00185
Roma, Italy.\\
$^2$ENEA, Centro Ricerche Frascati, U.T. Fus. (Fus. Mag. Lab.), Via Enrico Fermi 45, 00044 Frascati, Roma, Italy.\\
$^3$INFN (Istituto Nazionale di Fisica Nucleare), Sezione Roma1, Piazzale Aldo Moro 5, 00185 Roma, Italy
}
\date{November 2011}%

\begin{abstract}
The restriction to invariant connections in a Friedmann-Robertson-Walker space-time is discussed via the analysis of the Dirac brackets associated with the corresponding gauge fixing. This analysis allows us to establish the proper correspondence between reduced and un-reduced variables. In this respect, it is outlined how the holonomy-flux algebra coincides with the one of Loop Quantum Gravity if edges are parallel to simplicial vectors and the quantization of the model is performed via standard techniques by restricting admissible paths. Within this scheme, the discretization of the area spectrum is emphasized. Then, the role of the diffeomorphisms generator in reduced phase-space is investigated and it is clarified how it implements homogeneity on quantum states, which are defined over cubical knots. Finally, the perspectives for a consistent dynamical treatment are discussed. 
\end{abstract}

\pacs{04.60.Pp, 98.80.Qc}

\maketitle

\section{Introduction} 

The cosmological implementation of a quantum gravity model represents at the same time a physically relevant scenario, which describes the properties of the early Universe, and a privileged arena where the peculiarities of the full theory can be analyzed in a simplified framework. Loop Quantum Gravity (LQG) \cite{revloop} constitutes the most relevant approach in which the quantization of geometric degrees of freedom is performed. The definition of the kinematical Hilbert space\cite{kh}, the description of geometrical quantities in terms of quantum operators with discrete spectra \cite{ds,asharea} and the regularization of the super-Hamiltonian constraint \cite{th} are the most outstanding achievements of such a theory and they mark a close distinction with respect to the geometrodynamical Wheeler-DeWitt approach \cite{WdW}. However, the analysis of the super-Hamiltonian operator is such an hard task that no conclusion can be inferred for the semiclassical limit and the expected quantum modifications to Einstein equations. In this respect, the investigation of the full theory has been moved from the standard procedure to implement constraints {\it \'a-la} Dirac to different approaches, both in a canonical setting, via the so-called master constraint \cite{mc} and algebraic formulation \cite{af}, and in a covariant framework, through spin-foam models \cite{sf}. 

A proper dynamical description has been given in the framework of Loop Quantum Cosmology (LQC) \cite{lqc}, whose most impressive result is the avoidance of the initial singularity \cite{alqc}. Recently, the implications of LQC on the cosmic-microwave background radiation spectrum have been evaluated \cite{BCT}, so opening up the possibility to test experimentally quantum gravity corrections on the Universe dynamics.

These achievements have been made possible by the use of minisuperspace models, in which some degrees of freedom are frozen out and the dimensionality of the configuration space is reduced. For instance, the cosmological line element is described by the homogeneous and isotropic Friedman-Robertson-Walker (FRW) metric, in which there is an evolutionary variable only, the scale factor $a$. Henceforth, the reduced dynamical system is parametrized by a function of the scale factor and the whole space is represented by a single point on which $a$ lives. 

In LQC the minisuperspace approximation has been addressed by restricting SU(2) connections to invariant ones through some simplicial 1-forms \cite{mlqc}. This way, the reduced configuration space is parametrized by one time-dependent variable only, such that both the Gauss and diffeomorphisms constraints vanish identically. Hence, holonomies are defined by considering only those paths parallel to simplicial vectors (see \cite{sahal} for a discussion on this restriction) and quasi-periodic functions are taken as basic configuration variables. Finally, via the Gel'fand transform the Hilbert space is developed as the Bohr compactification of the real line.


The troubles come with the super-Hamiltonian operator $\mathcal{H}$. In fact, the standard picture to regularize $\mathcal{H}$ is based on fixing the minimum value for the parameter $\mu$ labeling quasi-periodic functions in such a way that the physical area spectrum has the same minimum as in LQG \cite{alqc}. In other word, the spatial manifold is enrolled down from the minisuperspace approximation. But if so, one cannot ignore that when considering the full spatial manifold, the Hilbert space is not anymore given by the Bohr compactification of the real line and the full SU(2)-graph structure must be considered. For instance in \cite{lqcfc}, we outlined that if one retains a local definition for reduced variables and consider the action of operators on SU(2) holonomies, the area operator has a discrete spectrum and the regularization of $\mathcal{H}$ in \cite{alqc} cannot be justified anymore. We proposed there a different approach to fix a minimum value for $\mu$, which is related with the total number of vertices of the fundamental path underlying the continuous space picture.

In this work, we are going to reformulate the steps which lead to define the variables describing a cosmological space-time, in order to achieve a quantum formulation closer to LQG with respect to the standard LQC framework. At first we will consider the gauge fixing of SU(2) invariance which provides the restriction to invariant triads, a part for some conformal space-dependent factors. The associated conditions will make the constraint hypersurfaces second-class and we will outline how the replacement of Poisson brackets with Dirac ones will allow us to define fluxes in reduced phase-space. In particular, the holonomy-flux algebra will be modified and it will reproduce the one of LQG only if holonomies will be evaluated along edges parallel to simplicial vectors. Henceforth, the quantization of the cosmological model will be performed according with the full theory and this would be possible in view of the local character that reduced variables retain after the gauge fixing. In particular, states will have support on reduced paths such that the quantum space can be described as a lattice with edges along simplicial vectors and the natural admissible vertices are 6-valent. 
We will consider the action of the area operator and outline that the associated spectrum is discrete. 

Furthermore, because the gauge fixing will not select invariant connections, but a dependence from spatial coordinates will remain in reduced variables, the super-momentum constraint will not vanish identically and we will analyze the role of the associated diffeomorphisms generator $D_a$. In particular, the vanishing of $D_a$ will imply the invariance under translations along simplicial vectors. By adopting the standard LQG tools to implement diffeomorphisms invariance on a quantum level, we will outline how translational invariance will impose homogeneity on a fundamental level.  
Then, given the form of the vertex structure as a 6-valent one, it will be discussed the possibility to regularize the super-Hamiltonian operator without any external assumption. Finally, we will investigate how the variables of LQC arise within this scheme and we will find the same duality condition between simplicial fluxes and curve length discussed in \cite{lqcfc}.  

The manuscript is organized as follows: in section \ref{2} LQG and LQC are reviewed and in section \ref{3} the SU(2) gauge fixing underlying the 
LQC formulation is discussed. Then, section \ref{4} is devoted to define the implications of such a gauge fixing in LQG kinematical Hilbert space, while in section \ref{5} a consistent way to impose the remaining diffeomorphisms invariance of the reduced phase-space is presented. In section \ref{6} the bridge with LQC phase-space is established via a proper reduction of configuration variables. Finally, brief concluding remarks follow in section \ref{7}.

\section{LQG and LQC} \label{2}

The main features of LQG are the reformulation of gravity in terms of a SU(2) gauge theory and the definition of the space of distributional connections. 
The former is made possible by the properties of gravity phase space \cite{prl}, which can be parametrized by the Ashtekar-Barbero connections $A^i_a$ and densitized triads $E^a_i$ as conjugate momenta. In particular, the smearing version of $A^i_a$ and $E^a_i$ (along a path $\Gamma(s)$ and across a surface $S$ with normal $n_a$, respectively) are taken as coordinates, such that basic Poisson brackets are described by the holonomy-flux algebra, {\it i.e.} (we work in units $8\pi G=\hbar=c=1$)\\\\
$[E_i(S),h_\Gamma]_{\mathrm{DB}}=$
\begin{eqnarray}
=\left\{\begin{array}{ccc} -i\gamma\sum_A o^{\Gamma,S}(s_A) h_\Gamma^{0,s_A}\tau_i h_\Gamma^{s_A,1} & \phantom1 & \Gamma\cap S=\{\Gamma^a(s_A)\}, \\
 0 & \phantom1 & otherwise \end{array}\right.,\label{hf}
\end{eqnarray}

where 
\begin{equation}
o^{\Gamma,S}(s_A)=\displaystyle\frac{n_a(s_A) (d\Gamma^a/ds)|_{s_A}}{|n_b(s_A) (d\Gamma^b/ds)|_{s_A}|},
\end{equation}

while $\gamma$ denotes the Immirzi parameter and $\tau_i$ are (Hermitian) SU(2) generators. 
 
The definition of distributional connections (see \cite{revloop} and references therein) is the basic point for a quantum representation which is inequivalent to the Wheeler-DeWitt one. The relationship with the classical configuration variables is established via SU(2) holonomies along $\Gamma$, which can be embedded into the space of general homomorphisms $X_{l(\Gamma)}$ from $\Gamma$ itself to the SU(2) group. The space of distributional connections $\bar{X}$ is made of homomorphisms from the set of all piecewise analytical paths of the spatial manifold (which can be seen as a tame subgroupoid \cite{vel}) into the topological SU(2) group. This space can be realized as the projective limit of $X_{l(\Gamma)}$, from which it can be demonstrated that $\bar{X}$ is a compact Hausdorff space in the Tychonov topology. This construction is particularly useful, because starting from the Haar measure of the SU(2) group, one can define the regular Borel probability measure $d\mu$ on $\bar{X}$. At the end, the kinematical Hilbert space can be introduced as the space of square integrable functions over $\bar{X}$, {\it i.e.} $\bar{A}=\mathcal{L}^2(\bar{X},d\mu)$, and essentially self-adjoint momentum operators can be implemented starting from the holonomy-flux algebra (\ref{hf}) \cite{ALMMT}. Thanks to the relation (\ref{hf}), the action of $E_i(S)$ can be described via the left invariant vector field of the SU(2) group. Basis vectors are invariant spin-networks, such that the quantum space is endowed with a combinatorial structure, given by the edges and the vertices of the fundamental paths underlying the continuous picture, which, together with the compactness of the gauge group, is responsible for the discreteness of geometrical operators spectra \cite{ds,asharea}.
 
LQC \cite{lqc} performs the quantization of a cosmological model via LQG-inspired techniques. The line element of a FRW space-time is given by 
\begin{equation}
ds^2 = dt^2-a^2\left(\frac{1}{1-kr^2}dr^2+d\Omega^2\right), \label{frw}
\end{equation}

$k$ being $1,0,-1$ for a closed, flat and open space, respectively. 

The starting point of such an approach consists in adopting invariant Ashtekar-Barbero connections \cite{bk} and densitized triads, which read as follows
\begin{equation}
A^i_a=c{}^0\!e^i_a,\qquad E^a_i=p{}^0\!e^a_i,\label{gf}
\end{equation} 

where $|p|={}^0\!e a^2$ and $c=\frac{1}{2}(k+\gamma \dot{a})$, while ${}^0\!e^a_i$ and ${}^0\!e^i_a$ denote the so-called simplicial 1-form and vectors, respectively, with ${}^0\!e=det({}^0\!e^a_i)$.

Hence, reduced phase-space is parametrized by $c$ and $p$, while holonomies are restricted to those edges parallel to simplicial vectors ${}^0\!e^i_a$, such that they take the following expression 
\begin{equation}
h_{i}=e^{i\mu c\tau_i}, \label{hlqc}
\end{equation}

$\mu$ being the edge length. Poisson brackets are given by
\begin{equation}
[p,h_i]_{\mathrm{PB}}=-i\mu\frac{\gamma}{3V_0}h_i\tau_i,\label{pb}
\end{equation}

$V_0$ being the volume of the fiducial metric. Holonomies (\ref{hlqc}) are determined by linear combinations of quasi-periodic functions $N_\mu=e^{i\mu c}$, thus $N_\mu$ themselves are taken as basis elements of the configuration space. The algebra generated by $\{N_\mu,p\}$ plays the role of the holonomy-flux algebra of LQG and by the analogous construction of the general case one finds that that $\bar{X}$ is the Bohr compactification of the real line $\textbf{R}_{Bohr}$. Hence, the Hilbert space is given by $\textsc{H}=\mathcal{L}^2(\textbf{R}_{Bohr},d\mu_{Bohr})$, where the measure reads 
\begin{equation}
<N_{\mu'}|N_\mu>=\delta_{\mu',\mu}.\label{dk}
\end{equation}

The momentum operator is implemented as essentially self-adjoint from the relation (\ref{pb}). 

Once the super-Hamiltonian operator is rewritten in terms of this set of variables, it takes the following expression in a particular operator ordering 
\begin{eqnarray} 
\mathcal{H}=\lim_{\bar\mu\rightarrow0}{H}^{\bar{\mu}}=\lim_{\bar\mu\rightarrow0}\bigg[sign(p)\frac{24i}{8\pi\gamma^3\bar{\mu}^3l_P^2}\sin{\bar{\mu}c}\nonumber\\\left(\sin{\frac{\bar{\mu}c}{2}}V\cos{\frac{\bar{\mu}c}{2}}-\cos{\frac{\bar{\mu}c}{2}}V\sin{\frac{\bar{\mu}c}{2}}\right)\sin{\bar{\mu}c}\bigg],\label{suph}
\end{eqnarray}

$\bar\mu$ being the length of edges entering the definition of $\mathcal{H}$, while $V=V_0p^{3/2}$. The limit above does not exists and $\mathcal{H}$ is evaluated by fixing a minimum $\bar\mu$ in such a way that the operator $p^2$ has the same minimum eigenvalue as the area operator of LQG. Such a description of a cosmological space-time leads to a difference equation, whose associated dynamics is not equivalent with the Wheeler-DeWitt one. The major achievement within this scheme is the replacing of the initial singularity with a bounce \cite{alqc}. 

In \cite{lqcfc} it has been outlined how the origin of the parameter $\bar\mu$ can be traced back to the combinatorial structure of the fundamental path underlying the continuous space ($\bar\mu=(V_0/N)^{1/3}$, $N$ being the total number of vertices), thus to the foundation of LQC.  

\section{SU(2) Gauge fixing} \label{3}

In LQC, SU(2) gauge invariance and the graph structure are lost, such that the tools adopted in LQG to infer discrete spectra for geometrical operators and to regularize the super-Hamiltonian are lost. We are now providing an alternative reduction of LQG to a cosmological space-time which is able to maintain these two properties and thus it can give a better perspective on a minisuperspace model for LQG.

In a FRW space-time one can perform an arbitrary spatial rotation, depending both on space and time coordinates, without changing the space-time geometry (\ref{frw}). This is the case because in the minisuperspace approximation the form of the metric is fixed and one is free to make arbitrary transformations in the tangent space. Hence, in order to work with invariant connections and triads (\ref{gf}) a fundamental gauge fixing of the SU(2) symmetry must be performed. In particular, the generic expression for $E^i_a$ can be obtained from (\ref{gf}) by a rotation, {\it i.e.}
\begin{equation} 
E^a_i=\sum_{j}p\Lambda^j_i {}^0\!e^a_j, 
\end{equation}     

$\Lambda^j_i$ being a generic SO(3) matrix, which is arbitrary as soon as gauge invariance is preserved. The condition $\Lambda^j_i=\delta^j_i$ is fixed by 
\begin{equation}
\chi_i=\epsilon_{ij}^{\phantom{12}k}{}^0\!e^j_a E^a_k=0.\label{chi}
\end{equation}

Indeed, the relation above is solved by 
\begin{equation}
E^a_i=p_i(x,t){}^0\!e^a_i,\label{pi}
\end{equation}

where we have three different $p_i$ which are functions of all space-time variables. At this level, let us choose 
\begin{equation}
p_1=p_2=p_3=p(x,t).\label{addcon}
\end{equation}

Hence, we have not performed yet the full reduction to invariant triads (\ref{gf}). However, the relation (\ref{chi}) fixes completely SU(2) gauge freedom. This can be seen from the fact that the Poisson brackets with the SU(2) Gauss constraint, {\it i.e.} 
\begin{equation}
G_i=\partial_aE^a_i+\gamma\epsilon_{ij}^{\phantom{12}k}A^j_aE^a_k, 
\end{equation}

do not vanish on the constraint hypersurfaces (\ref{pi}), but they give
\begin{equation}
[G_i,\chi_j]_{\mathrm{PB}}\approx -2\gamma {}^0\!e p\delta_{ij},
\end{equation}  

for which $det{[G_i,\chi_j]}\neq 0$. 

As a consequence, the full system of constraints is second-class and two different quantization procedures can be adopted: i) to work with reduced phase-space variables; ii)to describe the constraints hypersurfaces via the full set of phase-space coordinates, implementing non commuting conditions through the replacement of Poisson brackets with Dirac ones. 
 
LQC is based on i). Here we are going to fix the relationship between reduced and unreduced variables via the evaluation of Dirac brackets on the hypersurfaces (\ref{pi}) with the condition (\ref{addcon}).   

The Dirac brackets between unreduced phase-space variables are given by
\begin{eqnarray}
[A^i_a(x), A^j_b(y)]_{\mathrm{DB}}=\frac{1}{2}(E^i_{b}(y)A^j_{a}(x)-E^j_{a}(y)A^i_{b}(x)+\nonumber\\+2\delta^{ij}\delta_{kl}E^k_{[b}(y)A^l_{a]}(x))\delta^{(3)}(x-y)-\nonumber\\
-\frac{1}{2\gamma}\epsilon^{ij}_{\phantom{12}k}\bigg(E^k_{a}(x)\frac{\partial}{\partial x^{j}}\delta^{(3)}(x-y)+\nonumber\\+E^k_{b}(y)\frac{\partial}{\partial y^{a}}\delta^{(3)}(x-y)\bigg),
\end{eqnarray}
\begin{equation}
[A^i_a(x), E^b_j(y)]_{\mathrm{DB}}=\frac{1}{2}(\delta^i_j\delta_a^b+{}^0\!e^{bi}(x){}^0\!e_{aj}(y))\delta^{(3)}(x-y),\\
\end{equation}
\begin{equation}
[E^a_i(x), E^b_j(y)]_{\mathrm{DB}}=0.
\end{equation}

It is worth noting that connections do not commute anymore and this reflects the fact that on the constraint hypersurfaces the components $A^a_i$ are not independent. We can select as independent variable $c^i=\sum_{a}A^i_a{}^0\!e^a_i$, whose Dirac brackets are given by
\begin{equation}
[c^i(x,t),c^j(y,t)]_{\mathrm{DB}}=0,
\end{equation}
\begin{equation}
[c^i(x,t), E_j^b(y)]_{\mathrm{DB}}={}^0\!e_j^{b}(y)\delta^{i}_j\delta^{(3)}(x-y).\label{fdb}
\end{equation}

The first relation in the condition above ensures that $c^i$ are proper coordinates on the hypersurfaces $G_i=\chi_i=0$. These coordinates are precisely the set of variables on which the quantum formulation should be based on. In fact, the substitution of Poisson brackets with Dirac ones is not enough in order to perform a gauge fixing on a quantum level. One should also modify the scalar product in order to select proper coordinates on the constraints hypersurfaces \cite{ht}. 

The brackets (\ref{fdb}) will allow us to determine the action of un-reduced momenta after the gauge fixing, so fixing the spatial geometric structure of a cosmological space-time. The holonomies along a generic path $\Gamma$ evaluated on constraint hypersurfaces take the following expression
\begin{equation}
h_\Gamma=e^{i\sum_i\int_\Gamma c^i(x(s)){}^0\!e^i_a\frac{d\Gamma^a}{ds}ds\tau_i}.     
\end{equation}  

Hence, the holonomy-flux algebra after the gauge fixing reads\\\\ 
$[E_i(S),h_\Gamma]_{\mathrm{DB}}=$
\begin{eqnarray}
=\left\{\begin{array}{ccc} -i\gamma\sum_{A} \widetilde{o}^{\Gamma,S}_i(s_A) h_\Gamma^{0,s_A}\tau_i h_\Gamma^{s_A,1} & \phantom1 & \Gamma\cap S=\{\Gamma^i(s_A)\},\\
 \int C_i h_\Gamma(0,s)\tau_i h_\Gamma(s,1)ds & \phantom1 & \Gamma\subset S \\
 0 & \phantom1 & otherwise \end{array}\right.,\label{gfal}
\end{eqnarray}

where
\begin{equation}
\widetilde{o}^{\Gamma,S}_i(s_A)=\displaystyle\frac{n_a(s_A){}^0\!e^a_i(s_A) (d\Gamma^b/ds)|_{s_A}{}^0\!e_b^i(s_A)}{|n_c(s_A) (d\Gamma^c/ds)|_{s_A}|},
\end{equation}

while $C_i$ is a factor, which would need to be regularized. 

The expression (\ref{gfal}) gives the holonomy-flux algebra as soon as the gauge-fixing condition (\ref{chi}) holds. It is worth noting how the algebra undergoes a significant change, such that it is not possible to represent the action of fluxes in terms of left-invariant vector fields as in LQG. Furthermore, there is also a diverging contribution in the case when $\Gamma$ belongs to $S$. 

This issues can be avoided by restricting the class of paths to those ones along simplicial vectors, {\it i.e.}
\begin{equation}
\frac{d\Gamma^i}{ds}\propto{}^0\!e^i_c.
\end{equation}

In fact, in this case the factor $\widetilde{o}^{\Gamma,S}_i(s_A)$ coincides with the one of LQG $o^{\Gamma,S}(s_A)$, while $C_i=0$ for $\Gamma\in S$.    
In other words, the holonomy-flux algebra after the gauge fixing of SU(2) symmetry is still the one of the full theory if holonomies are restricted to those ones along edges parallel to simplicial vectors, which can be written as
\begin{equation}
h_{\Gamma^i}=e^{i\int c^i(s) ds\tau_i}.\label{0hol}
\end{equation}

In the following, we will consider those paths $\Gamma^i$ whose edges are straight along ${}^0\!e_a^i$ and we will denote them as reduced paths.

Therefore, the restriction to reduced paths can be motivated with the requirement to reproduce the holonomy-flux algebra of the full theory, which will allow us to define the action of variables associated with fluxes as essentially self-adjoint operators in the kinematical Hilbert space.

\section{Kinematical Hilbert space of gauge-fixed LQG}\label{4}
Because of the local character of reduced variables $c^i=c^i(x,t)$, the graph structure proper of LQG must be retained, the only restriction being that reduced paths must be considered. 
 
The definition of the space of distributional connections is made via the projective limit of $X_{l(\Gamma)}$, in which the paths enter merely as a label and the only requirement for them is that they form a partially oriented set. This means that arbitrary paths can be replaced with reduced ones and the definition of reduced distributional connections can be carried on as in the full theory.  Henceforth, the configuration space associated with a cosmological space-time can be developed as the projective limit $\bar{X}_C$ of homomorphisms from each reduced path $\Gamma$ to the SU(2) group and the associated Hilbert space is the one of square-integrable functions over $\bar{X}_C$, {\it i.e.} $\bar{A}_C=\mathcal{L}(\bar{X}_C,d\mu)$. Basis vectors are given by spin-network functionals defined on reduced paths only. 

The action of momenta operators follows from the expression (\ref{gfal}) and it is given by\\\\
$\hat{E}_i(S)h_{\Gamma^j}=$
\begin{eqnarray}
=\left\{\begin{array}{ccc} \gamma\sum_A \delta_{i}^jo^{\Gamma^jS}(s_A) h_{\Gamma^j}^{0,s_A}\tau_i h_{\Gamma^j}^{s_A,1} & \phantom1 & \Gamma^b\cap S=\{\Gamma^j(s_A)\},\\
 0 & \phantom1 & otherwise \end{array}\right.,\label{gfalq}
\end{eqnarray}

The relations above are crucial, because they demonstrate that the operators associated with $E^i_a$ are oriented along the simplicial 3-bein ${}^0\!E^i_a$. Therefore, only by taking properly into account the gauge fixing the resulting geometric structure reflects the one proper of simplicial vectors.  

The area operator of a surface $S$ intersecting $\Gamma$ in $s_A$ can be regularized as in \cite{asharea} and the final expression reads
\begin{equation}
\hat{A}[S]h_{\Gamma^i}=|\gamma|\sum_A n_a(s_A){}^0\!E^a_i(s_A) h_{\Gamma^i}|\tau_i|.
\end{equation}

This result emphasizes how the action of the area operator is determined by the SU(2) generators associated with the considered holonomy. In particular, $\hat{A}[S_i]$ is endowed with a discrete spectrum in view of the presence of $|\tau_i|$. 

\section{Relic 3-diffeomorphism invariance}\label{5}

The restriction to $E^a_i=p_i(x,t){}^0\!e^a_i$ and $A^i_a=c^i(x,t){}^0\!e^i_a$ implies that the supermomentum constraint does not vanish identically. In fact, there are some residual coordinate transformations which leave the fiducial metric invariant.
These transformations are generated by ${}^0\!e^i_a$ and we are going to discuss the kind of restriction they impose on a quantum level.       

In order to analyze the residual symmetries of a cosmological model, one must evaluate the action of the generator of 3-diffeomorphism with Dirac brackets, which coincide with working in reduced phase-space. In particular, one finds
\begin{eqnarray}
D[\vec{\xi}]=\int [\xi^a E^b_j\partial_aA^j_b-\xi^a \partial_b(A^i_aE^b_j)]d^3x=\nonumber\\=\int \sum_i[\xi^ap_i\partial_ac^i-\xi^a\partial_b(p_ic^i{}^0\!e^i_a {}^0\!e^b_i)]d^3x,
\end{eqnarray}

$\xi^a$ being arbitrary parameters.

The brackets between $p_i$ and $c^j$ can be inferred from $p_i=E^a_i{}^0\!e^i_a$, such that one can evaluate the expression above finding
\begin{equation}
[D[\vec{\xi}],c^j(x)]=[\xi^a\partial_ac^j+c^j{}^0\!e^j_a {}^0\!e^b_j \partial_b\xi^a]|_x, 
\end{equation}

and the finite action on reduced holonomies is such the following
\begin{equation}
[D[\vec\xi],h_{\Gamma^i}]=h_{\phi(\Gamma^i)}-h_{\Gamma^i},
\end{equation}

where the path $\phi(\Gamma^i)$ is obtained from $\Gamma^i$ by the translation $x^i\rightarrow x^i-\xi^a{}^0\!e^i_a$. In other words, in the phase-space parametrized by holonomies over reduced paths, full 3-diffeomorphisms invariance reduces to the invariance under translations along simplicial vectors. This feature can be implemented as in LQG, by taking functionals in the dual space defined over ``reduced knots'', {\it i.e.} the equivalence class of reduced paths under translations.

Let us now consider the action of finite diffeomorphisms in the kinematical Hilbert space. Each reduced path is the direct product of 1-dimensional paths along each simplicial direction. 1-dimensional paths over the same reduced graph are a partition of the same line, on which edge base-points define a lattice. Admissible paths are associated with all the possible lattices one can realize. Let us denote by $P^I=\{e^I_A\}$ a certain partition of the line. A generic state is the direct sum over each partition $P^I$ of reduced spin networks along edges $e^I_A$, {\it i.e.}   
\begin{equation}
\psi=\oplus_{I}d_I\otimes_A \sum_jc_{A,I,j} T^j_{e^I_A},
\end{equation} 

$c_{A,I}$, $d_I$ being generic coefficients, while $T^j_{e^I_A}$ denote the reduced j-spin-network along the edge $e^I_A$. The action of a finite smooth diffeomorphisms maps a partition $P^I$ into a partition $P^J$, $I\neq J$, such that the total number of edges is preserved. This way, a solution can be developed in the dual space by taking a sum over the orbit. This means that physical states do not depend on which particular partition is chosen. The resulting picture is that of cubical knot. 

However, there also exists a class of diffeomorphisms preserving a generic partition $I$, which are those translations mapping one-to-one all edges $e^I_A$ into $e^I_B$ with $B\neq A$. The invariance under this class of transformations implies that $c_{A,I,j}=c_{B,I,j}$. In other words, the quantum state associated with any given edge $A$, {\it i.e.} $\sum_jc_{A,I,j} T^j_{e^I_A}$, must be the same along the whole partition $I$.     

As a consequence, the emerging spatial manifold exhibits a fundamental homogeneity along each direction. This is due to the fact that the whole diffeomorphisms group has been reduced to the invariance under translations along each simplicial direction, which act on 1-dimensional paths only. 

In order to implement isotropy, the additional condition (\ref{addcon}) must be imposed. This can be done simply by identifying the quantum states along different directions, so realizing the cosmological principle on a quantum level. 

What remain to be done is the description of 3-dimensional vertices via a proper reduction of invariant intertwiners $i^{mnpqrs}$. In fact,
given the vertex structure, the expression of the un-reduced super-Hamiltonian operator can be regularized according with the procedure described in \cite{th} for the full theory, the difference being that reduced paths have to be considered only. Such a procedure is based on introducing a path-dependent triangulation of the spatial manifold and on defining $\mathcal{H}$ as an operator acting at vertices. The resulting expression is automatically regularized without any external assumption, as the imposition of a minimum value for the parameter $\bar\mu$ in LQC. Moreover, the fixed 6-valent vertex structure is expected to provide a sensible simplification of the analytic treatment, such that the issues of the full theory can be overcome. 

Therefore, the description of a cosmological space in terms of a cubical knot with a fixed quantum state attached to each edge will allow us to discuss a self-consistent cosmological implementation of LQG and to investigate the foundation of LQC.

\section{Fundamental LQC variables}\label{6}

The configuration space of LQC is derived from holonomies (\ref{0hol}) as soon one imposes that $c^i$ does not depend on spatial coordinates, such that
\begin{equation}  
h_{\Gamma^i}=e^{i\mu \overline{c}_\Gamma\tau_i},  
\end{equation}

where $\overline{c}_\Gamma$ can be written as $\overline{c}_\Gamma=\int_\Gamma cdt/\mu$. Once the independence from spatial coordinates is assumed, Dirac brackets with 3-bein densitized vectors (\ref{fdb}) becomes
\begin{equation}
[\overline{c}_\Gamma, E^j_b(y)]_{\mathrm{DB}}=\left\{\begin{array}{cc} \frac{1}{\mu}{}^0\!E^j_{b}(y)\delta^{(2)}(\Gamma(s)-y), & y=\Gamma(s) \\
0 & y\notin\Gamma \end{array}\right..
\end{equation}

The resulting holonomy-flux algebra is still the same as in LQG and the discretization of the area operator spectrum follows as discussed in the section \ref{4}. 

The variable $p$ can be defined from the action of fluxes (\ref{gfalq}) and a possible definition is the following one 
\begin{equation}  
p=\frac{1}{3}\sum_{i=1}^3p_i,\qquad p_i=\hat{E}_i(S_i)/\Delta, 
\end{equation}

$S_i$ being the surface whose normal vector is ${}^0\!e^i_a$, while $\Delta$ is the associated flux in the simplicial metric (we take the same $\Delta$ in each direction). This way, the Dirac brackets between $p$ and holonomies gives
\begin{equation} 
[p,h_{\Gamma^i}]_{\mathrm{DB}}=-i\frac{\gamma}{3\Delta}h_{\Gamma^i}\tau_i.
\end{equation}

In order to reconcile the expression above with the analogous one in LQC (\ref{pb}) one has to fix $\Delta\mu=V_0$, so finding the same duality condition as in \cite{lqcfc}. 

Therefore, the length of curves and the flux across surfaces in the simplicial manifold are related and the parameter $\mu$ does not enter the spectrum of the area operator. According with the results of \cite{lqcfc}, this point suggests that the regularization of the reduced super-Hamiltonian operator can be performed fixing $\bar\mu=(V_0/N)^{1/3}$, $N$ being the total number of vertices. Henceforth, as far as LQC variables are identified, the origin of the minimum $\bar\mu$ value can be traced back to the existence of a fundamental path underlying the continuous picture and, in particular, to the number of vertices of such a path. In this respect, it is worth noting that in the previous section we found that the number of edges, thus of vertices, labeled different superselection sectors under the action of reduced diffeomorphisms.

\section{Conclusions}\label{7}
In this paper, we investigated the foundation of LQC and the fate of SU(2) degrees of freedom in the associated gauge-fixed formulation (\ref{gf}). We analyzed the Dirac brackets emerging by the restriction to constraint hypersurfaces and the non-commutativity of connections has been pointed out. Then, we determined the holonomy-flux algebra in the gauge-fixed formulation and we outlined that it coincided with the one of the full theory as far as the restriction to edges parallel to simplicial vectors took place. Henceforth, we could quantize the whole system as in LQG but on a special class of paths, characterized by a 6-valent vertex structure. As a consequence, the area operator was defined and the discreteness of the corresponding spectrum has been emphasized. 

The diffeomorphisms group reduced to translations along simplicial vectors. By working with reduced paths, we could solve the associated constraints by using the same techniques adopted in full LQG. In particular, we concluded that quantum states in the physical Hilbert space were defined on the direct product of the equivalence classes of reduced paths under translations. This fact implied that along each simplicial direction all edges were associated with the same quantum state. This way, homogeneity had been imposed on the path structure. As soon as the restriction to FRW variables took place, the resulting picture of the fundamental cosmological space was that of a homogeneous and isotropic cubical knot.

Finally, the relationship with the standard LQC formulation has been inferred by defining the operator $p$ from fundamental fluxes. This analysis provided the same duality condition between simplicial fluxes and curve length as in \cite{lqcfc} and it posed serious doubts on the viability of the super-Hamiltonian regularization procedure proper of LQC \cite{alqc}. 

However, here the possibility to describe a cosmological line-element by a path having all edges parallel to simplicial vectors and a fixed vertex structure opens up the perspective to address the dynamical behavior directly from the action of the super-Hamiltonian operator regularized as in LQG  \cite{th}. This point is currently under investigation \cite{al}, because it is expected to provide non-trivial results on the quantum evolution of a cosmological space-time and, at the same level, it can give us hints on the dynamics of the full model.  

\section{Acknowledgment}   

The work of F.C. has been supported in part by an exchange visit grant funded by the European Science Foundation, received in the framework of the Research Networking Programme on ``Quantum Geometry and Quantum Gravity''.

\end{document}